\documentclass[11pt]{article}

\usepackage[round]{natbib}
\usepackage{graphicx}
\usepackage[margin=1in]{geometry}
\usepackage{url}
\usepackage{bibentry}

\usepackage{amsmath}
\usepackage[linesnumbered,ruled,algo2e]{algorithm2e}
\usepackage{mathabx}
\usepackage[hidelinks]{hyperref}

\newcommand\blfootnote[1]{%
	\begingroup
	\renewcommand\thefootnote{}\footnote{#1}%
	\addtocounter{footnote}{-1}%
	\endgroup
}

\newcommand\fscore{\mathop{\mbox{$F$-$\mathit{score}$}}}

\title{How do people describe locations during a natural disaster: an analysis of tweets from Hurricane Harvey} 

\author{
	Yingjie Hu and Jimin Wang
	\\GeoAI Lab, Department of Geography, University at Buffalo, NY, USA  
}
\date{}

\begin{document}

\maketitle

\begin{abstract}
Social media platforms, such as Twitter, have been increasingly used by people during natural disasters to share information and request for help. Hurricane Harvey was a category 4 hurricane that devastated  Houston, Texas, USA in August 2017 and caused catastrophic flooding in the Houston metropolitan area. Hurricane Harvey also witnessed the widespread use of social media by the general public in response to this major disaster, and geographic locations are key information pieces  described in many of the social media messages. A  geoparsing system, or a geoparser, can be utilized to automatically extract and locate the described locations, which can help first responders reach the people in need. While a number of geoparsers have already been developed, it is unclear how effective they are in recognizing and geo-locating the locations described by people during natural disasters. To fill this gap, this work seeks to understand how people describe locations during a natural disaster by analyzing a sample of tweets posted during Hurricane Harvey. We then identify the limitations of existing geoparsers in processing these tweets, and discuss possible approaches to overcoming these limitations.

\vspace{0.2cm}
\end{abstract}

\section{Introduction}
Hurricane Harvey was a Category 4 tropical storm which started on
August 17, 2017 and ended on September 2, 2017 and made a
landfall on Texas and Louisiana, USA. It dropped more than 1,300 mm of rain over the Houston metropolitan area and caused catastrophic flooding \cite[]{zhang2018urbanization}. During the hurricane and the subsequent flooding, social media platforms, such as Twitter, were  used by many residents in the city of Houston and the surrounding areas  to share disaster-related information and send help requests. \blfootnote{\hspace{-0.6cm}*This is a preprint. The final version is published as: \vspace{0.1cm}\\ 
 Hu, Y. and Wang, J. (2020): How Do People Describe Locations During a Natural Disaster: An Analysis of Tweets from Hurricane Harvey. In K. Janowicz and J. A. Verstegen, editors, the 11th International Conference on Geographic Information Science (GIScience 2021) - Part I, volume 177, pages 6:1-6:16, Dagstuhl, Germany. DOI: \url{https://doi.org/10.4230/LIPIcs.GIScience.2021.I.6}}

The use of social media during natural disasters is not new. An early work by  \cite{de2009omg}  used Twitter to analyze a  forest fire in the South of France back in July 2009. In the following years, many studies were conducted based on the social media data collected from   disasters to understand the emergency situations on the ground and  the reactions  of the general public. Examples include  the 2010 Pakistan flood \cite[]{murthy2013twitter}, the 2011 earthquake on the East Coast of the US \cite[]{crooks2013earthquake}, Hurricane Sandy in 2012 \cite[]{middleton2013real}, the 2014 wildfire of California  \cite[]{wang2016spatial},  Hurricane Joaquin in 2015 \cite[]{wang2018hyper}, and Hurricane Irma in 2017 \cite[]{yu2019deep}. Social media data, such as tweets, provide near real-time information about what is happening in the disaster-affected area, and  are suitable for applications in disaster response and situational awareness \cite[]{maceachren2011senseplace2}.  Twitter, in particular, allows researchers to retrieve about 1\% of the total number of public tweets for free via its API, and this ability enables various tweet-based disaster studies.

While social media has already been used  in disasters and emergency situations,  Hurricane Harvey was probably the first major disaster in which the use of social media  was comparable or even surpassed the use of some traditional communication methods during a disaster. The National Public Radio (NPR) of the US published an article with the headline  “Facebook, Twitter Replace 911 Calls For Stranded In Houston” \cite[]{silverman2017facebook}, which described how  social media platforms were widely used by Houston residents  to request for help when  911 could not be reached. The fact that the storm took out over a dozen emergency call centers and that there were too many 911 calls during and after the hurricane were among the reasons responsible for the  failure of the 911 system. Another article published in The Wall Street Journal was titled ``Hurricane Harvey Victims Turn to Social Media for Assistance'', which described similar stories in which people turned to social media for help after their 911 calls  failed \cite[]{seetharaman2017harvey}. In addition, Hurricane Harvey was called by The Time Magazine  as ``The U.S.'s First Social Media Storm'' \cite[]{rhodan2017harvey}. Besides news articles, a  survey was conducted by researchers \cite[]{mihunov2020use} after Hurricane Harvey, which filtered through 2,082 people  in Houston and the surrounding
communities, and focused on 195 Twitter users. They found that about one-third of their respondents indicated that they used social media to request for help because they were unable to connect to 911. 

With the ubiquity of smart mobile devices and the  popularity of social media, it seems to be a natural choice for people to turn to Twitter, Facebook, or other social media platforms when their 911 calls fail. People are already familiar with the basic use of these social media platforms (e.g., how to create a post and how to upload a photo), and they can stay connected with their friends and family members online, follow the latest information from public figures (e.g., the Twitter account of the mayor of the affected city), authoritative agencies (e.g., FEMA), and voluntary organizations, and can ``@'' related people and organizations to send targeted messages. Indeed, a survey by  \cite{pourebrahim2019understanding} based on Hurricane Sandy in 2012  revealed that Twitter users received emergency information faster and from more sources than non-Twitter users. The survey by  \cite{mihunov2020use} found that about 76\% of their respondents considered Twitter as ``very useful'' or ``extremely useful''   for seeking help during Hurricane Harvey, and  roughly three quarters of their respondents indicated that Twitter and other social media were easy to use. Their survey also revealed some challenges in the use of Twitter during a natural disaster, such as not knowing whether volunteers received their requests or when they would send help. However, these situations could change in future disasters, as volunteers and relief organizations learn to collect the   requests  from social media. In addition to Twitter, other social media platforms were also used by people to seek help \cite[]{li2019using}. For example, an online group named “Hurricane Harvey 2017 - Together We Will Make It” was created on Facebook  to enable  victims to post messages about their situations during the flooding \cite[]{silverman2017facebook}.

One major challenge in handling the help requests that people sent on social media platforms is to efficiently process the huge number of posts. As described by a disaster responding consultant during Hurricane Harvey \cite[]{silverman2017facebook}, ``It is literally trying to drink from a firehose''. Disaster responders simply do not have the bandwidth and time to manually monitor the huge number of posts on social media and identify actionable information. In fact, there exist multiple challenges in effectively using the information from social media platforms, including verifying the veracity of the posted information, understanding the purpose of the posts (e.g., whether a post is about requesting rescue, reporting disaster situation, calling for donation, or praying for the affected people), and extracting critical information pieces (e.g., the locations of the people who need help). Much research has already been devoted to identifying true information from false information \cite[]{gupta2013faking,vosoughi2018spread}, classifying the purposes of social media posts \cite[]{imran2014aidr,ashktorab2014tweedr}, and extracting information from tweets \cite[]{imran2013extracting,ragini2018big}.

This paper focuses on the specific challenge of extracting locations  from the tweets posted during a natural disaster. 
As a first step, we focus on understanding how people describe locations during a disaster by analyzing a sample of tweets randomly selected from over 7 million tweets posted during Hurricane Harvey. The contribution of this paper is twofold:
\begin{itemize}
	\item We conduct an  analysis  on a sample of 1,000 randomly selected tweets to understand and categorize the ways people describe locations during a natural disaster.
	\item We identify the limitations of existing tools in extracting locations from these tweets and discuss possible approaches to overcoming these limitations.
\end{itemize}

The remainder of this paper is organized as follows. Section 2 reviews related work in geoparsing and tweet analysis in the context of disasters. Section 3 describes the dataset from Hurricane Harvey. In Section 4, we analyze and classify location descriptions in the selected tweets. Section 5 reports the experiment results of using existing tools for processing the tweets. Finally, Section 6 summarizes this work and discusses future directions.



\section{Related work}
\begin{figure}[h]
	\centering
	\includegraphics[width=0.95\textwidth]{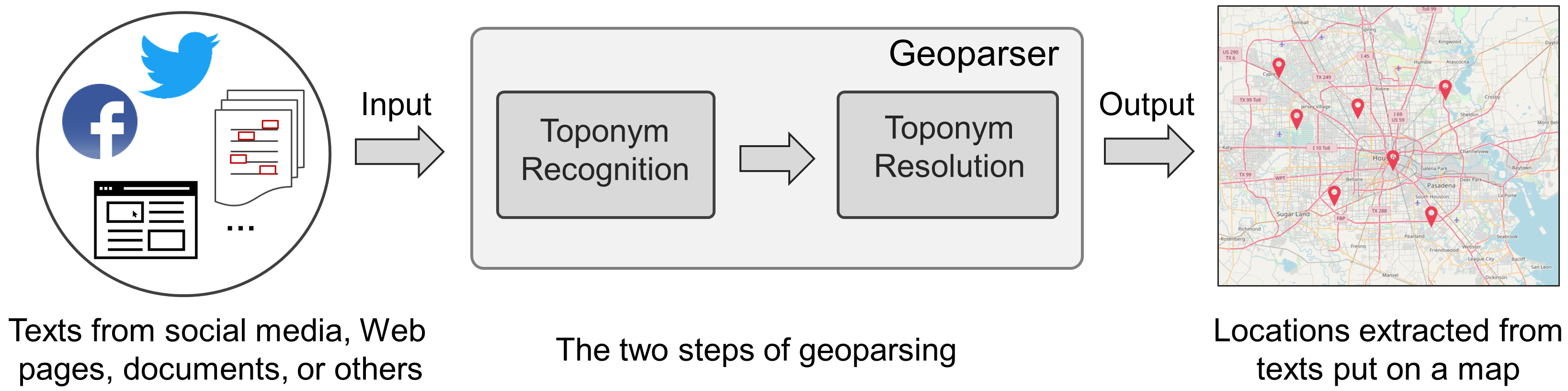}
	\caption{The typical process of geoparsing  text to extract locations.}\label{overview}
		\vspace*{-0.2cm}
\end{figure}

Locations in tweets can be extracted through  \textit{geoparsing},  a process of recognizing and geo-locating place names (or toponyms) from texts \cite[]{freire2011metadata,gritta2018s,wang2019enhancing}. Geoparsing is often studied within the  field of geographic information retrieval (GIR) \cite[]{doi:10.1080/13658810701626343,purves2018geographic}. A software tool developed for geoparsing is called a \textit{geoparser}, which typically functions in two consecutive steps: \textit{toponym recognition} and \textit{toponym resolution}. The first step recognizes toponyms from texts, and the second step resolves any place name ambiguity and assigns suitable geographic coordinates. Figure \ref{overview} illustrates these two steps. It is worth noting that geoparsing can be applied to other types of texts in addition to social media messages, such as Web pages, news articles, organization documents, and others.

A number of geoparsers have already been developed by researchers. GeoTxt is an online geoparser developed by \cite{karimzadeh2013geotxt,karimzadeh2019geotxt},  which uses the Stanford Named Entity Recognition (NER) tool and several other NER tools for toponym recognition and employs the GeoNames gazetteer\footnote{https://www.geonames.org/} for toponym resolution. TopoCluster, developed by  \cite{delozier2015gazetteer}, is a geoparser that  uses the Stanford NER for  toponym recognition and leverages a technique based on the geographic word profiles  for toponym resolution. The Edinburgh Geoparser,  developed by the Language Technology Group at Edinburgh University \cite[]{alex2015adapting}, uses their own natural language processing (NLP)  tool, called LT-TTT2, for toponym recognition, and  a gazetteer (e.g., GeoNames) and pre-defined heuristics for toponym resolution. Cartographic Location And Vicinity INdexer (CLAVIN)\footnote{https://clavin.bericotechnologies.com/Berico\_CLAVIN.pdf} is a geoparser developed by Berico Technologies that employs the NER tool from the Apache OpenNLP library or the Stanford NER for toponym recognition, and utilizes a gazetteer and heuristics for toponym resolution. CamCoder is a  toponym resolution model developed by  \cite{gritta2018melbourne}, which integrates a convolutional neural network and geographic vector representations. Gritta et al. further converted CamCoder  into a geoparser by employing the  spaCy NER tool  for  toponym recognition.

Twitter data were used in many previous studies on  situational awareness and disaster response. 
 \cite{imran2014aidr} and  \cite{yu2019deep}  developed machine learning and text mining systems for automatically classifying  tweets into topics, e.g., \textit{caution and advice} and \textit{casualty and damage}.  \cite{huang2015geographic} classified tweets into different disaster phases, such as preparedness, response, impact and recovery.  \cite{kryvasheyeu2016rapid} and  \cite{li2018novel} used tweets for assessing the damages of disasters. Existing studies, however, often used only the geotagged locations of tweets \cite[]{de2015geographic,wang2016spatial} or the locations in the profiles of Twitter users  \cite[]{zou2018mining,zou2019social}, rather than the locations described in tweet content. Many geotagged locations  were collected by the GPS receivers in smart mobile devices, and therefore are generally more accurate than the locations geoparsed from the content of tweets. This can be a reason that motivated researchers to use the geotagged locations of tweets. Meanwhile, geotagged locations reflect only the current locations of Twitter users, which may not be the same as the locations  described in the content of tweets. In addition, only about 1\%  tweets were geotagged \cite[]{sloan2013knowing}, and the number of geotagged tweets further decreased with Twitter's removal of precise geotagging in June 2019. By contrast, researchers found that over 10\% tweets contain some location references in their content \cite[]{maceachren2011senseplace2}. For the locations in the profiles of Twitter users, they may reflect neither the current locations of the users nor the locations described by the users, since the profile locations can be their birthplaces, work places, marriage places, or even imaginary places, and are not always updated. 

Some research  examined location extraction  from the content of tweets.  GeoTxt is a geoparser originally developed for processing tweets \cite[]{karimzadeh2019geotxt}; however, their testing experiments were based on a tweet corpus, GeoCopora \cite[]{geocorpora2018}, whose toponyms are mostly country names and major city names, rather than fine-grained place names in a disaster affected area (although GeoCopora does contain some fine-grained locations, such as school names).  \cite{gelernter2013algorithm}  geoparsed locations in the  tweets from the 2011 earthquake in Christchurch, New Zealand, and  \cite{wang2018hyper}  extracted locations from tweets for monitoring the flood during Hurricane Joaquin in 2015. However, both work focused on using a mixture of NLP techniques and packages (e.g., abbreviation expansion, spell correction, and NER tools) for location extraction, rather than a more detailed analysis on the characteristics of the location descriptions. This paper aims to fill such a gap by examining how people describe locations  in tweets  during a natural disaster, with the ultimate goal of helping design more effective geoparsers for assisting disaster response.

\section{Dataset}
The dataset used in this work is a set of 7,041,866 tweets collected during Hurricane Harvey and the subsequent flooding from August 18, 2017 to September 22, 2017. This dataset was prepared by the University of North Texas Libraries, and the  tweets  were retrieved based on a set of hashtags and keywords, such as “\#HurricaneHarvey”, “\#HoustonFlood”, and ``Hurricane Harvey''. The entire dataset is available from the library repository of North Texas University (NTU)\footnote{https://digital.library.unt.edu/ark:/67531/metadc993940/}, and it is in the public domain. 

\begin{figure}[h]
	\centering
	\includegraphics[width=\textwidth]{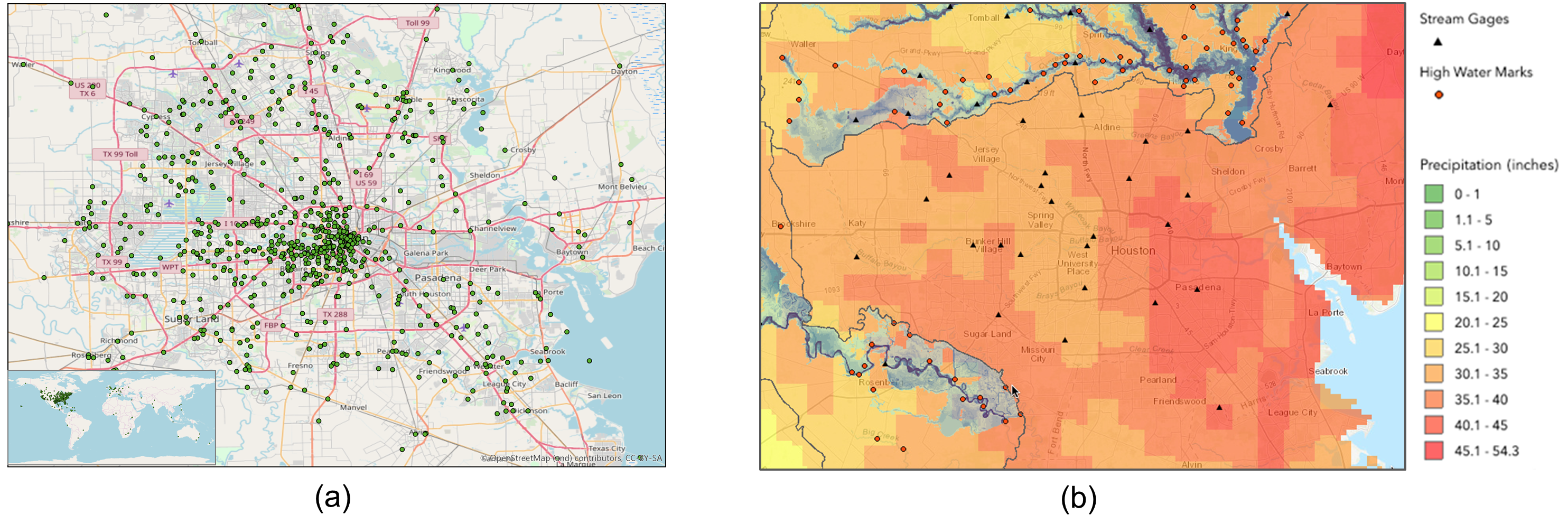}
	\caption{A comparison of the locations of geotagged tweets and the precipitation during Hurricane Harvey: (a) locations of the geotagged tweets; (b) precipitation in the Houston area from the USGS.}\label{tweets}
	\vspace*{-0.2cm}
\end{figure}

Among the over seven million tweets in the entire dataset, only 7,540 are geotagged with longitude and latitude coordinates. These geotagged tweets are distributed not only  within the Houston area but also throughout the world, with most of the tweets located inside the United States. Figure \ref{tweets}(a) shows the locations of the geotagged tweets in the Houston area, and the locations of all the geotagged tweets are visualized in the overview map in the lower-left corner. 
The low percentage of geotagged tweets (about 0.1\%) in this dataset and the fact that the geotagged tweets are distributed throughout the world can be attributed to the data collection process:  the data were collected using a list of keywords and hashtags rather than focusing on a particular geographic area. 
We  compare the locations of the geotagged tweets with the precipitation map\footnote{https://webapps.usgs.gov/harvey/} from the US Geological Survey (USGS) (Figure \ref{tweets}(b)). No clear relationship can be visually identified between the locations of the geotagged tweets and the severity of the precipitation in different areas. For example, the northwestern region received relatively less precipitation than the southeastern region, but there were more geotagged tweets in the former region.

In this work, we are particularly interested in the locations described in the content of tweets. While both the news  and literature told us that people  used Twitter and other social media platforms to request for help and share information, we still do not know how specifically people describe locations in social media messages during this natural disaster. Manually analyzing the 7,041,866 tweets is  practically impossible. Thus, we use a simple regular expression to narrow down the target tweets to be analyzed. The regular expression contains about 70  location-related terms that are frequently observed in place names and location descriptions, such as ``street'', ``avenue'', ``park'', ``square'', ``bridge'', ``rd'', and ``ave''. A full list of these terms and the constructed regular expression can be accessed at: \url{https://github.com/geoai-lab/HowDoPeopleDescribeLocations}.  Running this regular expression against the 7 million tweets returns  15,834 tweets. A quick examination of these  15,834 tweets shows that many of them contain detailed location descriptions, such as  house number addresses or   school names.  For curiosity, we also run the same regular expression against the 7,540 geotagged tweets. Only 203 tweets are returned. This result suggests that there are many tweets that contain location descriptions but are not geotagged. Thus, we will miss  important information if we focus on geotagged locations only. 

We randomly select 1,000 tweets from the 15,834 records returned by the regular expression. This  selection is performed as follows:  we first remove retweets to avoid duplication; we then index the remaining tweets, and  generate 1,000  non-repeating random integers that are used  as the indexes to retrieve the corresponding tweets. 
As we read through some of these tweets, we see vivid images of people seeking help and sharing information during Hurricane Harvey. Some examples are provided as below:
\begin{itemize}
	\item \textit{``12 Y/O BOY NEEDs RESCUED! 8100 Cypresswood Dr Spring TX 77379 They are trapped on second story! \#houstonflood''}
	\item \textit{``80 people stranded in a church!!  5547 Cavalcade St, Houston, TX 77026 \#harveyrescue  \#hurricaneharvey''}
	\item \textit{``Rescue needed: 2907 Trinity Drive, Pearland, Tx. Need boat rescue 3 people, 2 elderly one is 90 not steady in her feet \& cant swim. \#Harvey''}
	\item \textit{``Community is responding at shelters in College Park High School and Magnolia High School \#TheWoodlands \#Harvey…''}
	\item \textit{``\#Houston \#HoustonFlood the intersection of I-45 \& N. Main Street''}
\end{itemize}   

While the above  tweets certainly do not represent all of those posted during Hurricane Harvey, they demonstrate the urgency of some requests. Effectively and efficiently extracting locations from these  tweets can help responders and volunteers to reach the people at risk more quickly and can even save lives. In addition, these examples  also show that some people were  requesting help for others. Thus, even if their tweets were geotagged, it is necessary to focus on the locations described in the content  rather than the geotagged locations. 









\section{Understanding the locations described in Harvey tweets}
In this section, we examine and understand the ways people describe locations based on the 1,000  tweets. To do so, we carefully read through each of the tweets, identify and annotate the locations described in their content, and classify the location descriptions. It is worth noting that  we focus on the descriptions that refer to specific geographic locations rather than general \textit{locative expressions} \cite[]{liu2014automatic}, such as ``this corner'' or ``that building''. The data annotation is done in the following steps. First, the second author  reads each tweet and annotates the location descriptions identified; second, the first author goes through the entire  dataset, checking each location annotation and discussing with the second author to resolve any annotation difference; a preliminary list of location categories is also identified in this step; third, the first author   goes through the entire dataset again, refines the list of categories, and classifies the location descriptions; fourth, the second author performs another round of checking to examine the classified location descriptions. The locations are annotated using the IOB model widely adopted in the CoNLL shared tasks \cite[]{tjong2003introduction}. 
In the process of annotating the data, we also find that some of the initial 1,000 tweets  do not contain specific locations (e.g., a tweet may say: ``My side street is now a rushing tributary''). We replace those tweets with others randomly selected from the rest of the data, so that each of the 1,000 tweets contains at least one specific location description. The annotated dataset is available at: \url{https://github.com/geoai-lab/HowDoPeopleDescribeLocations}. 

Ten categories of location descriptions are identified based on the 1,000 Hurricane Harvey tweets (Table \ref{tab:category}). The number of tweets in each category is also summarized in Table \ref{tab:category} in the column \textit{Count}. It is worth noting that a tweet may contain more than one type of location descriptions, and therefore can be counted toward more than one category. 
\begin{table}[!ht]
		\caption{Ten categories of location descriptions identified from the 1,000 Harvey tweets.}
		\vspace*{0.2cm}
	\begin{tabular}{|l|l|l|}
		\hline
		\textit{\textbf{Category}}                                                            & \textit{\textbf{Examples}}                                                             &
		\textit{\textbf{Count}}  
		                                                                                             \\ \hline
		\begin{tabular}[c]{@{}l@{}}C1: House \\ number addresses\end{tabular}                 & \textit{\begin{tabular}[c]{@{}l@{}}- "Papa stranded in home. Water rising above waist. HELP\\ \textbf{8111} \textbf{Woodlyn}      \textbf{Rd, 77028} \#houstonflood"\\ - "\#HurricaneHarvey family needs rescuing at \textbf{11800}\\ \textbf{Grant Rd. Apt. 1009.}     \textbf{Cypress, Texas 77429}"\end{tabular}}            &   257                                                                        \\ \hline
		C2: Street names                                                                      & \textit{\begin{tabular}[c]{@{}l@{}}- "\#Harvey LIVE from San Antonio, TX. Fatal car \\accident at \textbf{Ingram Rd.},    Strong winds."\\ - "\textbf{Allen Parkway}, \textbf{Memorial}, \textbf{Waugh overpass}, Spotts \\ park and Buffalo     Bayou park completely under water"\end{tabular}}                            &   571                                                           \\ \hline
		C3: Highways                                                                          & \textit{\begin{tabular}[c]{@{}l@{}}- "9:00AM update video from Hogan St over White Oak \\  Bayou, \textbf{I-10},   \textbf{I-45}: water down about 4' since last night…"\\ - "Left Corpus bout to be in San Angelo \#HurricaneHarvey\\Y'all be safe  Avoided \textbf{highway 37}  Took the back road"\end{tabular}}                                   &   68                                   \\ \hline
		\begin{tabular}[c]{@{}l@{}}C4: Exits of \\        highways\end{tabular}               & \textit{\begin{tabular}[c]{@{}l@{}}- "Need trailers/trucks to move dogs from Park Location: \\  Whites Park    Pavillion off \textbf{I-10 exit 61} Anahuac TX"\\ - "\textbf{TX 249 Northbound} at Chasewood Dr.  \textbf{Louetta Rd.}\\    \textbf{ Exit}. \#houstonflood"\end{tabular}}              &   8                                                                \\ \hline
		\begin{tabular}[c]{@{}l@{}}C5: Intersections \\ of roads (rivers)     \end{tabular}               & \textit{\begin{tabular}[c]{@{}l@{}}- "Guys, this is \textbf{I-45} at \textbf{Main Street} in Houston. Crazy.  \\    \#hurricane \#harvey…"\\ - "Major flooding at \textbf{Clay Rd} \& \textbf{Queenston} in west Houston.\\Lots    of rescues going on for ppl trapped..."\end{tabular}}                      &   109                                                 \\ \hline
		\begin{tabular}[c]{@{}l@{}}C6: Natural \\        features\end{tabular}                & \textit{\begin{tabular}[c]{@{}l@{}}- "\textbf{Buffalo Bayou} holding steady at 10,000 cfs at the gage\\ near Terry Hershey Park"\\ - "Frontage Rd at the river \#hurricaneHarvey \\\#hurricaneharvey    @ \textbf{San Jacinto River}"\end{tabular}}                          &   77                                                            \\ \hline
		\begin{tabular}[c]{@{}l@{}}C7: Other\\human-made        \\features\\        \end{tabular} & \textit{\begin{tabular}[c]{@{}l@{}}- "Houston's \textbf{Buffalo Bayou Park} - always among the first\\     to flood. \#Harvey"\\ - "If you need a place to escape \#HurricaneHarvey, The \\ \textbf{Willie De}   \textbf{Leon Civic Center}: 300 E. Main St in\\ Uvalde is open as a shelter"\end{tabular}}                             &   219                                         \\ \hline
		\begin{tabular}[c]{@{}l@{}}C8: Local \\       organizations\end{tabular}              & \textit{\begin{tabular}[c]{@{}l@{}}- "\#Harvey does anyone know about the flooding conditions\\  around    \textbf{Cypress Ridge High School}?! \#HurricaneHarvey"\\ - "Cleaning supply drive is underway. 9-11 am today\\at \textbf{Preston Hollow Presbyterian Church}"\end{tabular}}              &   60                \\ \hline
		\begin{tabular}[c]{@{}l@{}}C9: Admin\\        units\end{tabular}                     & \textit{\begin{tabular}[c]{@{}l@{}}- "\#HurricaneHarvey INTENSE eye wall of category 4\\ Hurricane    Harvey from \textbf{Rockport, TX}"\\ - "Pictures of downed trees and damaged apartment building \\ on Airline   Road in \textbf{Corpus Christi}."\end{tabular}}                                           &   644                   \\ \hline
		\begin{tabular}[c]{@{}l@{}}C10: Multiple\\        areas\end{tabular}               & \textit{\begin{tabular}[c]{@{}l@{}}- "\#HurricaneHarvey Anyone doing high water rescues in the \\    \textbf{Pasadena/Deer Park} area? My daughter has been stranded \\in a parking  lot all night"\\ - "FYI to any of you in \textbf{NW Houston/Lakewood Forest},\\ Projections   are showing Cypress Creek overflowing at Grant Rd"\end{tabular}}  &   6  \\ \hline
	\end{tabular}\label{tab:category}
\vspace*{-0.3cm}
\end{table}

For  category  \textit{C1}, we are surprised to see many tweets using the very standard U.S. postal office address format, with a house number, street name, city name, state name, and postal code. Those house number addresses, once effectively extracted from the text, can be located via a typical geocoder (although today's geocoders and geoparsers are developed as separate tools). Some addresses  only contain a house number and a street name. Those addresses can  be  located by narrowing down to the area that is affected by the disaster, e.g., Houston or Texas in the case of Hurricane Harvey.

Categories \textit{C2} and \textit{C3} cover location descriptions about roads and highways. These two categories could be merged into one. We separate them because our experiments later find that existing NER tools have difficulty in recognizing the US highway names, such as \textit{I-45} and \textit{Hwy 90}. Yet, those highway names are  common in many geographic areas of the US and in the daily conversations of people. Thus, we believe that this category is worth to be highlighted from the perspective of developing better geoparsers. 

Category \textit{C4}  covers highway exits.  People can use an exit to provide a more precise location related to a highway. They may use the exit number, e.g., \textit{``exit 61''}, or  the street name of an exit, e.g., \textit{``Louetta Rd. Exit''}.  This may be related to the US culture since  road signs on the US highways often provide both the exit numbers and the corresponding street names. 
One may also use two exits in one tweet to describe a segment of a highway, such as in \textit{``My uncle is stuck in his truck on I-45 between Cypress Hill \& Huffmeister exits''}.

Category  \textit{C5} covers location descriptions related to road (or river) intersections. We identify five ways  used by people in tweets to describe road intersections: (1) Road A and Road B, (2) Road A \& Road B, (3) Road A at Road B,  (4) Road A @ Road B,  (5) Road A / Road B. Besides, people often use abbreviations when describing intersections, e.g., they may write \textit{``Mary Bates and Concho St''} instead of  \textit{``Mary Bates Blvd and Concho St''}. The intersections of two rivers, or a road and a river,  are described in  similar ways, such as in \textit{``White Oak Bayou at Houston Avenue 1:00 pm Saturday \#Houston''}. A tweet may contain more than one intersection, such as in \textit{``Streets Flooded: Almeda Genoa Rd. from Windmill Lakes Blvd. to Rowlett Rd.''} which uses two  intersections to describe a road segment.

Categories \textit{C6}, \textit{C7}, and \textit{C8} cover location descriptions related to natural features, other human-made features (excluding streets and highways), and local organizations. These location descriptions are generally  in the form of place names, such as the name of a bayou, a church, a school, or a park. We find that many tweets also provide the exact  address in addition to a place name, such as the second example of \textit{C7}. 

Category \textit{C9} covers location descriptions related to towns, cities, and states. Examples include \textit{Houston}, \textit{Katy}, \textit{Rockport}, \textit{Corpus Christi}, \textit{Texas}, and \textit{TX}. This type of locations has limited value from a disaster response perspective, due to their coarse geospatial resolutions.

Category \textit{C10} covers locations related to multiple areas.  We find that people use this way to describe a geographic region  that typically involves two smaller neighborhoods, towns, or cities, such as \textit{``Pasadena''} and \textit{``Deer Park''} in the first example. 

In summary, we have identified ten categories of location descriptions based on the 1,000 tweets from Hurricane Harvey. 
Overall, people seem to describe their locations precisely by  providing the exact house number addresses, road intersections, exit numbers of highways, or adding detailed address information to place names. One  reason may be that people, when under  emergency situations, may choose to describe locations in precise ways   in order to be understood by  others such as first responders and volunteers. 
While these categories are identified based on the 1,000 tweets from a particular disaster, they seem to be general and are likely to be used by people in future disasters in the US. Understanding these location descriptions is fundamental for designing  effective geoparsers to support disaster response.










\section{Extracting locations from  Harvey tweets using existing tools}
With the 1,000 Harvey tweets annotated, we examine the performance of existing tools on extracting locations from these tweets. While this seems to be a straightforward task, there are limitations in existing geoparsers  that prevent their direct application. First, none of the five geoparsers that we  discussed previously, namely GeoTxt, TopoCluster, CLAVIN, the Edinburgh Geoparser, and CamCoder, have the capability of geocoding house number addresses which are an important type of location descriptions  (the category of \textit{C1}). Second, none of the five geoparsers have the capability of geo-locating local street names and highway names (the categories of \textit{C2} and \textit{C3}) at a sub-city level (largely due to their use of the GeoNames gazetteer which focuses on the names of cities and upper-level administrative units), let alone road intersections and highway exits (the categories of \textit{C4} and \textit{C5}). It is worth noting that these limitations do not suggest that existing geoparsers are not well designed; instead, they suggest that there is a gap between the demand of processing disaster-related tweets focusing on a local area and the expected application of the existing geoparsers  for extracting city- and upper-level toponyms throughout the world (the category of \textit{C9}). Such an application fits well with one of the important objectives of GIR research, namely to geographically index documents such as Web pages \cite[]{amitay2004web}. 
Although we cannot directly apply existing geoparsers to the Harvey tweets, we can examine their components on toponym recognition and resolution respectively.

\subsection{Toponym recognition}
Existing geoparsers typically use off-the-shelf NER tools for the step of toponym recognition rather than designing their own models. A rationale of doing so is that toponym recognition, to some extent, can be considered as a subtask of named entity recognition. 
Indeed, many NER tools can recognize multiple types of entities from text, such as persons, companies, locations, genes, music albums, and others. Thus, one can use an  NER tool for toponym recognition by keeping only \textit{locations} in the output, and save the effort of developing a model from scratch. How would the  NER tools used in existing geoparsers perform on  the Hurricane Harvey tweets? In the following, we conduct experiments to answer this question.

The NER tools to be tested in our experiments are the Stanford NER and the spaCy NER, both of which are used in existing geoparsers. Particularly, the Stanford NER  has been used in GeoTxt, TopoCluster, and CLAVIN, and the spaCy NER has been used in CamCoder. The Stanford NER has both a default version, which is sensitive to upper and lower letter cases, and a caseless version. Considering that the content of tweets may not have regular capitalization as in well-formatted text, we test both the default case-sensitive Stanford NER and the caseless version. With the typically used 3-class model, both case-sensitive and caseless Stanford NER have three classes in their output:  \textit{Person}, \textit{Organization}, and \textit{Location}. Given the names of the three classes, one might  choose to keep \textit{Location} only in the output. However, doing so will miss schools and churches described in the tweets, which are often used as shelters during a disaster, because the Stanford NER considers schools and churches as \textit{Organization}. An alternative choice is to keep both \textit{Location} and \textit{Organization} in the output. However, such a design choice will include false positives. For example, in the sentence ``The Red Cross has provided recovery assistance to more than 46,000 households affected by Hurricane Harvey'', \textit{``Red Cross''} will be included in the output since it is an \textit{Organization}; this adds a false positive into the toponym recognition result. 
The spaCy NER has a similar issue, whose output includes multiple classes related to geography. These classes are \textit{Facility} (e.g., buildings, airports, and highways), \textit{Organization} (e.g., companies, agencies, and institutions), \textit{GPE} (Geo-Political Entity; e.g., countries, cities, and states), and \textit{Location} (e.g., non-GPE locations, mountain ranges, and bodies of water). Again, one might  choose to keep \textit{Location} only given the names of these classes, and a direct consequence is that the spaCy NER will only recognize natural features, such as rivers and mountains, and will miss all other valid toponyms. On the other hand, keeping all  the classes can introduce  false positives into the output of the spaCy NER. In this work, we test these different design choices for the Stanford NER and the spaCy NER. Specifically, we examine the performances of the Stanford NER when only \textit{Location} is kept in the output (we call it \textit{``narrow''} version) and when  both \textit{Organization} and \textit{Location} are kept (\textit{``broad''} version). For the spaCy NER, we examine its performances when only \textit{Location} is kept (\textit{``narrow''}) and when all location-related entities are kept (\textit{``broad''}). In total, we test six versions of the NER tools: the default Stanford NER (narrow and broad), the caseless Stanford NER (narrow and broad), and the spaCy NER (narrow and broad).

In the first set of experiments, we evaluate the  performances of these NER tools on recognizing all locations regardless of their categories from the 1,000 Hurricane Harvey tweets. The metrics used   are \textit{precision}, \textit{recall}, and \textit{F-score} (Equations 1-3). 
\begin{align}
  \hspace*{3cm}  Precision &= \frac{tp}{tp + fp} \\
 \hspace*{3cm}   Recall &= \frac{tp}{tp + fn} \\
 \hspace*{3cm}   \fscore &= 2\cdot\: \frac{Precision \times Recall}{Precision + Recall}
\end{align} 
\textit{Precision} measures the percentage of correctly recognized locations  (true positives or \textit{tp}) among all the locations that are recognized by the model (both \textit{tp} and false positives (\textit{fp})). \textit{Recall} measures the percentage of correctly recognized locations among all the annotated locations  which include \textit{tp} and false negatives (\textit{fn}). \textit{F-score} is the harmonic mean of the precision and the recall. F-score is high only when both precision and recall are fairly high, and is low if either of the two is low. 

The performances of the six versions of NER tools are reported in Table \ref{tab:performance}.
\begin{table}[!htbp]
	
	\caption{Performances of the NER tools on the 1,000 Hurricane Harvey tweets.}
		\vspace*{0.2cm}
	\centering
	\begin{tabular}{|l|c|c|c|}
		\hline
		\textit{\textbf{NER tool}}                                           & \textit{\textbf{Precision}} & \textit{\textbf{Recall}} & \textit{\textbf{F-score}} \\ \hline
		\begin{tabular}[c]{@{}l@{}}Stanford default\\ (\textit{Narrow})\end{tabular}  & \textbf{0.829}                       & 0.400                    & 0.540                     \\ \hline
		\begin{tabular}[c]{@{}l@{}}Stanford default\\ (\textit{Broad})\end{tabular}   & 0.733                       & \textbf{0.441}                    & \textbf{0.551}                     \\ \hline
		\begin{tabular}[c]{@{}l@{}}Stanford caseless\\ (\textit{Narrow})\end{tabular} & 0.804                       & 0.321                    & 0.458                     \\ \hline
		\begin{tabular}[c]{@{}l@{}}Stanford caseless\\ (\textit{Broad})\end{tabular}  & 0.723                       & 0.337                    & 0.460                     \\ \hline
		\begin{tabular}[c]{@{}l@{}}spaCy NER\\ (\textit{Narrow})\end{tabular}             & 0.575                       & 0.024                    & 0.046                     \\ \hline
		\begin{tabular}[c]{@{}l@{}}spaCy NER\\ (\textit{Broad})\end{tabular}              & 0.463                       & 0.305                    & 0.367                     \\ \hline
	\end{tabular} \label{tab:performance}
\end{table}
Overall, the performances of all four versions of the Stanford NER dominate the spaCy NER. This result suggests the effectiveness of this classic and open source NER tool developed by the Stanford Natural Language Processing Group \cite[]{manning2014stanford}. The default Stanford NER with a \textit{narrow} output (i.e., keeping \textit{Location} only) achieves the highest precision, while the default Stanford NER with a \textit{broad} output (i.e., keeping both \textit{Location} and \textit{Organization}) achieves the highest recall and F-score. This result can be explained by the capability of the \textit{broad} Stanford NER in recognizing schools, churches, and other organizations that are also locations in these Hurricane Harvey tweets. The lower precision of the \textit{broad} Stanford NER compared with the \textit{narrow} Stanford NER is explained by the included false positives of the \textit{broad} version. Another interesting observation  from the result is that the default Stanford NER overall performs better than the caseless Stanford NER. Since tweets are user-generated content that may not  follow the regular upper and lower cases, we may be tempted to use the caseless version of the Stanford NER.  While there do exist tweets with ill-capitalized words, we find that a large percentage of the analyzed tweets (over 85\%) still use correct  capitalization. Thus, using a caseless version of the Stanford NER, which completely ignores  letter cases  in the text, will miss the  information contained in the correct capitalization used by  many tweets. On the other hand, if one expects that most capitalization in the text is incorrect or the text is not capitalized at all, then the caseless  version is  likely to be a better choice.

In the second set of experiments, we evaluate the performances of the NER tools on the different categories of location descriptions reported in Table \ref{tab:category}. Here, we cannot use the same \textit{Precision}, \textit{Recall}, and \textit{F-score} as the evaluation metrics. This is because these NER tools do not differentiate the  ten categories of locations (e.g., the Stanford NER considers all of the entities as \textit{Location} or \textit{Organization}, while the spaCy NER does not differentiate streets, highways, and other human-made features). Thus, we use the metric of \textit{Accuracy} that has been used in previous studies, such as \cite[]{gelernter2011geo,karimzadeh2016performance,gritta2018s,wang2019enhancing}.  It is calculated using the equation below:
\begin{equation}
\hspace*{3cm} Accuracy_c = \frac{|Recognized \cap Annotated_c |}{ |Annotated_c|}
\end{equation}
where $Accuracy_c$ represents the \textit{Accuracy} of a model on the location category $c$;  $Recognized$ represents the set of all locations recognized by the model; and $Annotated_c$ represents the set of annotated locations in the category $c$. 

\begin{figure}[h]
	\centering
	\includegraphics[width=0.95\textwidth]{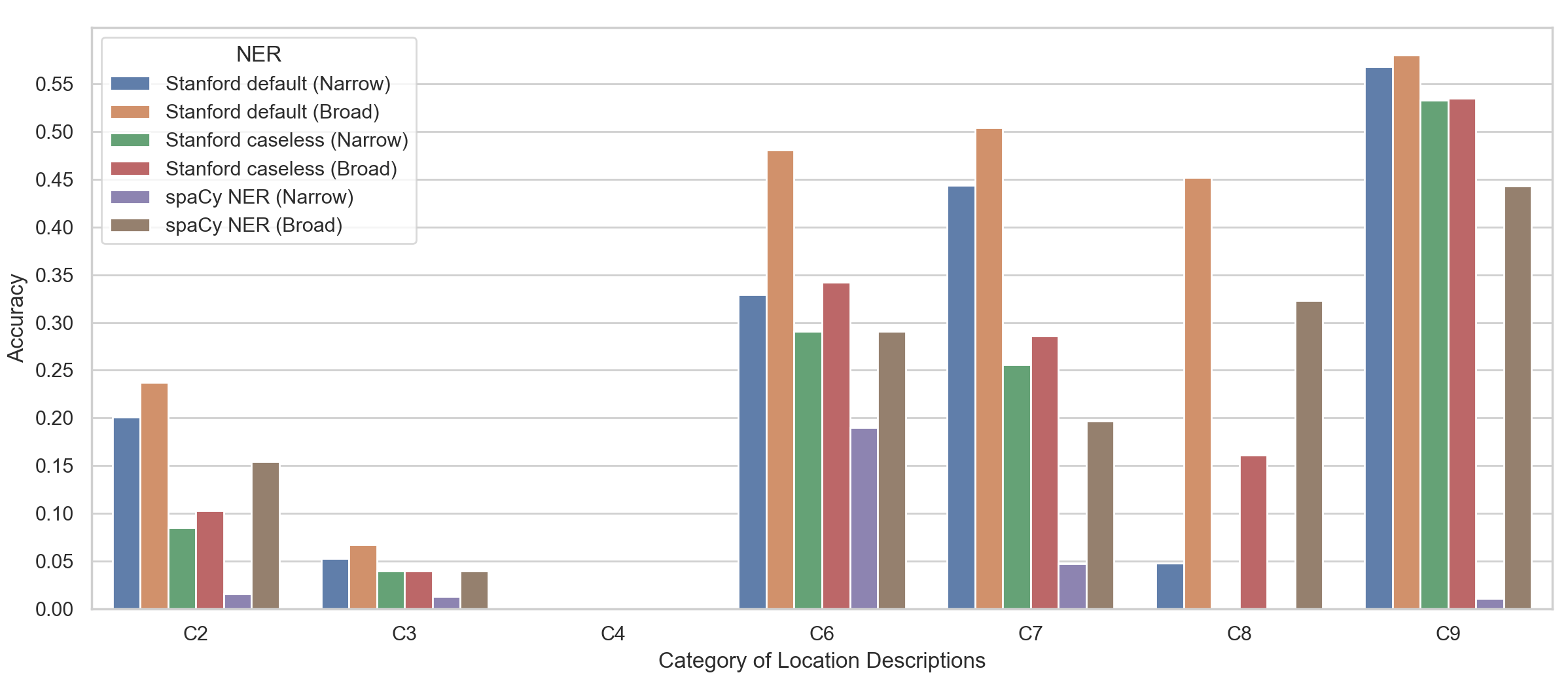}
	\caption{Performances of the NER tools on the different categories of location descriptions.}\label{performance}
	\vspace*{-0.2cm}
\end{figure}

In addition, an NER tool cannot recognize a location that consists of multiple entities. For example, a house number address like \textit{``5547 Cavalcade St, Houston, TX 77026''} (category \textit{C1}) consists of a door number, a street name, a city name, a state name, and a zip code, which are typically recognized as separate entities by an NER. Similar situation applies to road intersections (category \textit{C5}) and multiple areas (category \textit{C10}). These three categories are thus not included in the experiments. The performances of the NER tools on the other seven location categories are shown in Figure \ref{performance}.

A number of observations can be obtained from the result. First, all six versions of the NER tools fail on the category \textit{C4: Exits of highways}. This suggests a major limitation of using these off-the-shelf NER tools for toponym recognition: they will miss all the rescue requests whose locations are in the form of highway exits. Second, the broad version of the default Stanford NER has the highest \textit{accuracy} across different categories of location descriptions. However, the broad version likely sacrifices \textit{precision} for \textit{recall} (which cannot be directly measured for each individual category), given its lower  overall \textit{precision}  compared with the narrow version  reported in Table \ref{tab:performance}. As can be seen in Figure \ref{performance}, the broad version of the Stanford NER shows a major gain in recognizing organizations (\textit{C8}), since it includes entities in the type of \textit{Organization} in the output. While the broad version also recognizes more locations in other categories, this is often because those locations are considered as \textit{Organization} by the Stanford NER in general. For example, centers, such as \textit{``Walnut Hill Rec Center''} and \textit{``Delco Center''}, in our category \textit{C7} are considered as \textit{Organization} by the Stanford NER. Third, five out of the six NER tools recognize fair percentages of administrative unit names (the category of \textit{C9}), such as   \textit{``Houston''} and \textit{``Texas''}. The only exception is the narrow version of the spaCy NER, since it only recognizes the names of natural geographic features. Despite the fair performances of the NER tools, this category of locations  has limited value for disaster rescue purposes. Fourth, the performances of the NER tools on street names (\textit{C2}) and highway names (\textit{C2}) are low, but these location description are usually critical for locating the people who need help.  A more detailed examination of the result shows that these NER tools often miss the street names that contain numbers, such as \textit{26th St} and \textit{31st Ave}.  Similarly, they miss  the highway names, such as \textit{I-10} and \textit{Hwy 90}, in which numbers are used even more frequently than in street names. Finally,  these NER tools have only low to fair performances on natural features (e.g., rivers and bayous; \textit{C6}) and other human-made features (e.g., parks; \textit{C7}). 

In sum, the experiment results suggest that existing NER tools have limited performance in recognizing locations, especially sub-city level locations, from disaster-related tweets. They do not have the capability of recognizing location descriptions that consist of multiple entities, such as house number addresses, road intersections, and multiple areas, and largely fail on highways, highway exits, and the street names that contain numbers.
As a result, there is a need for  developing more effective toponym recognition models that can recognize these location descriptions from  tweets.

\subsection{Toponym resolution}
The toponym resolution components of existing geoparsers use a variety of strategies to resolve ambiguity and geo-locate place names. These strategies include heuristics based on the population of cities (e.g., a toponym is  resolved to the place with the highest population), the co-occurrences of related place names  (e.g., the names of  higher administrative units), and others \cite[]{alex2015adapting,karimzadeh2019geotxt}. There are also methods that create a grid tessellation covering the surface of the Earth and calculate the probability of a place name to be located in each grid \cite[]{delozier2015gazetteer,gritta2018melbourne}. However, existing toponym resolution components focus more on the task of disambiguating and geo-locating place names at a world scale, such as understanding which \textit{``Washington''} the place name is referring to, given the many places named \textit{``Washington''} in the world. 

By contrast, the task of resolving locations described in disaster related tweets has  different characteristics. First, these locations are generally at sub-city level, such as roads and house number addresses. Unlike cities, these fine-grained locations are often not associated with populations. This makes it difficult to apply existing toponym resolution heuristics based on population. 
Second, given these location descriptions are about a disaster-affected local area,  the task of toponym disambiguation  becomes easier. While there can still be roads having the same name within the same city, the number of places that share the same name decreases largely (e.g., there is no need to disambiguate over 80 different \textit{``Washington''}s when we focus on a local area). Third, point-based location representations typically returned by existing geoparsers become insufficient. We may need lines or polygons, in addition to points, to provide more accurate representation for the described locations. 

Given that  existing toponym resolution strategies are not applicable to the task of  resolving location descriptions in disaster-related tweets, we discuss what are needed if we are going to develop a toponym resolution model for handling this task. First, it is necessary to have a local gazetteer that focuses on the disaster affected area and has detailed geometric representation (i.e., points, lines, and polygons) of the geographic features. Compared with the typically used GeoNames gazetteer,  a local gazetteer serves two roles: (1) it reduces place name ambiguity  by limiting place names  to the disaster-affected area; and (2) it provides detailed spatial footprints for representing  fine-grained locations. Such a local gazetteer could be constructed by conflating OpenStreetMap data, the GeoNames data within the local region, and  authoritative geospatial data from mapping agencies. Second, we need a geocoder embedded in the toponym resolution model to handle house number addresses. Successfully embedding such a geocoder also requires the local gazetteer to contain house number data along with the roads and streets. Third, additional natural language processing methods are necessary to identify the spatial relations among the multiple locations described in the same tweet. This is especially important for location descriptions in Categories \textit{C4}, \textit{C5}, and \textit{C10} when we need to locate the intersection of two roads (or a road and a river), the exit of a highway, or a combination of two regions. In addition, the NLP methods can  help the toponym resolution model determine which geometric representation to use. Consider two possible tweets \textit{``Both Allen Parkway and Memorial Dr are flooded''} and \textit{``Flooding at the intersection of  Allen Parkway and Memorial Dr''}. While the same  roads are described in these two tweets, the ideal geometric representation for them should be different.





\section{Conclusions and future work}
Hurricane Harvey is a major natural disaster that devastated the Houston metropolitan area in 2017. Hurricane Harvey also witnessed the wide use of social media, such as Twitter, by the disaster-affected people to seek help and share information. Given the increasing popularity of social media among the general public, they are likely to be used in future disasters.  One challenge in using  social media messages for supporting disaster response is automatically and accurately extracting locations from these messages. In this work, we examine a sample of tweets sent out during Hurricane Harvey in order to understand how people describe locations in the context of a natural disaster. We identify ten categories of location descriptions, ranging from house number addresses and highway exits to human-made features and multiple regions. We find that under emergency situations people tend to describe their locations  precisely by providing  exact house numbers or clear road intersection information. We further conduct experiments to measure the performances of existing tools for geoparsing these Harvey tweets. Limitations of these  tools are identified, and we discuss possible approaches to developing more effective models. In addition to social media messages, other approaches, such as \textit{what3words} (\url{what3words.com}), could also be promoted to help people communicate their locations in emergency situations. \textit{What3words} could be especially useful in geographic areas that lack standard addresses; meanwhile, it will also require people to have some familiarity with the system and install the relevant app.


A number of research topics can be pursued in the near future. First, while we have gained some understanding on how people describe locations during a natural disaster, it is limited to English language and within the culture of the United States. People speaking other languages or in other countries and cultures are likely to describe locations in different ways that need further investigation. Second, we can move forward and experiment  possible approaches to developing models for recognizing and geo-locating the location descriptions in tweets posted during disasters. Examples include  toponym recognition models that can correctly recognize highways and streets whose names contain numbers, and toponym resolution models that can correctly interpret the spatial relations of the multiple locations described in the same tweet. Finally, location extraction is only one part (although an important part) of the whole pipeline for deriving useful information from social media messages. Future research can integrate location extraction with other methods, such as those for verifying information veracity and classifying message purposes, to help disaster responders and volunteer organizations make more effective use of social media and reach the people in need.



\bibliographystyle{plainnat}
\bibliography{lipics-v2019-sample-article}

\end{document}